\documentclass[12pt]{article}
\usepackage{amssymb} 
\usepackage{amsfonts} 
\usepackage{amsthm}
\usepackage{latexsym} 
\newtheorem{prop}{Proposition}

\newtheorem{lemma}{Lemma} 
\theoremstyle{definition}
\newtheorem*{defn}{Definition} 
\newcommand{\abs}[1]{\mbox{$|#1|$}}

\newcommand{\norm}[1]{\mbox{$\|#1\|$}} 

\newcommand{\alg}[1]{\mbox{$\mathcal{#1}$}}
\newcommand{\hil}[1]{\mbox{$\mathcal{#1}$}}
\newcommand{\inte}[1]{\mathrm{int}(#1)}
\newcommand{\clo}[1]{\mathrm{clo}(#1)}
\newcommand{\quo}{\mathcal{B}(\mathbb{R})/\mathcal{N}}
 
\newcommand{\df}{:=}
\newcommand{\bor}[1]{\mathcal{B}(#1)} 
\newcommand{\comp}{\gamma \mathbb{R}} 
\newcommand{\buc}{BUC(\mathbb{R})}
\newcommand{\val}[1]{\mathrm{Val}(#1)}
\newcommand{\bz}{\mathbf{0}}

\title{On the nature of continuous physical quantities in classical and 
quantum mechanics }
\author{Hans Halvorson\footnote{Department of Philosophy, University
    of Pittsburgh; Email: hphst1+@pitt.edu }}
\begin{document}
\maketitle

\begin{abstract}  Within the traditional Hilbert space formalism of
  quantum mechanics, it is not possible to describe a particle as
  possessing, simultaneously, a sharp position value and a sharp
  momentum value.  Is it possible, though, to describe a particle as
  possessing just a sharp position value (or just a sharp momentum
  value)?  Some, such as Teller~\cite{teller}, have thought that the
  answer to this question is No --- that the status of individual
  continuous quantities is very different in quantum mechanics than in
  classical mechanics.  On the contrary, I shall show that the same
  subtle issues arise with respect to continuous quantities in
  classical and quantum mechanics; and that it is, after all, possible
  to describe a particle as possessing a sharp position value without
  altering the standard formalism of quantum mechanics.
\end{abstract}

\section{Introduction}
Consider the following experimental setup: We have a source of
quantum-mechanical particles (e.g. electrons), a fluorescent screen,
and a barrier between the source and the screen which has one or more
openings.  The barrier, with its openings, prepares an ensemble of
particles whose state, upon arrival at the screen, is given by some
wavefunction $\psi$.  Let's assume for simplicity that the screen is
one-dimensional and infinite in both directions (i.e., represented by
$\mathbb{R}$).  The screen should be thought of as a measuring device
for the position of particles in the ensemble, and we know that the
pattern of shading on the screen will be distributed in accordance
with the squared modulus $\abs{\psi}^{2}$ of the wavefunction.  For
any individual particle in the ensemble, and any region $S$ on the
screen, $\int _{S}\abs{\psi}^{2}$ gives the probability that the the
screen will flash inside $S$ when the particle hits the screen.  It
would be very natural then to attempt to interpret the probability
distribution $\abs{\psi}^{2}$ as a measure of our ignorance about the
precise state of individual particles in the ensemble.  That is, it is
natural to think that each particle in the ensemble does in fact have
a ``sure-fire'' (probability $1$) disposition to be found at a certain
point $\lambda \in \mathbb{R}$, and probabilities arise out of our
ignorance about precisely which point $\lambda$ that is.

It is clear that the traditional interpretive obstacles in quantum
theory do not prevent us from thinking this way of this measurement.
For one, since we are not asking how a particle gets from the source
to the screen, interference effects play no role.  Second, since we
are in a position measurement context (i.e., we are ignoring other
observables such as momentum), there can be no question of the
no-hidden-variables theorems presenting any obstacles to our thinking
of $\abs{\psi}^{2}$ as a mixture of pure position states.

However, there remains a serious difficulty for supplying an ignorance
interpretation of $\abs{\psi}^{2}$: There does not appear to be any
object in the standard Hilbert space formalism of QM which could
represent the state of an individual particle in the ensemble.
According to Teller~\cite{teller}, then, attempting to give an
ignorance intepretation to $\abs{\psi}^{2}$ amounts to questioning the
``descriptive completeness'' of quantum theory:
\begin{quote} ``...if imprecise or imperfectly defined values of a
  continuous quantity are to be explained in terms of assumed
  underlying states with exact point values, these assumed underlying
  states cannot be described within the theory, and the theory is
  descriptively incomplete.''  (p. 347) \end{quote} But since QM is
descriptively complete, $\abs{\psi}^{2}$ should not be interpreted as
a measure of our ignorance of the precise state of the individual
particles.

Now it is certainly true that no \emph{vector} in Hilbert space can be
taken to describe the state of an individual particle with a sharp
location.  This is just the familiar point that operators with
continuous spectra, like position, have no eigenvectors.  Furthermore,
Gleason's Theorem~\cite{gle} entails that any (pure) $\sigma$-additive
probability measure on the logic of quantum propositions (i.e.,
lattice of subspaces of a Hilbert space) is given by a vector $\psi$,
via the standard formula $E\mapsto \langle \psi |E\psi \rangle$.  I
believe, however, that the $\sigma$-additivity assumption in Gleason's
theorem is too restrictive.  Indeed, I will argue that if we required
$\sigma$-additivity of the states of a \emph{classical} mechanical
system, then we would similarly have to conclude --- in direct
contradiction to what we know to be true --- that there can be no
states assigning sharp values to continuous quantities in that case as
well.  In fact, I will show that in both cases it is possible to
assign sharp values to continuous quantities using finitely-additive
(``singular'') states.  I will also demonstrate that difficulties
which arise when using singular states in conjunction with the
traditional von-Neumann account of ideal measurement are resolved when
we adopt a more realistic account of measurement.
 
\section{Observables, properties, and states}
It is well-known that there are methods of extending the Hilbert space
formalism of QM so that it includes ``generalized eigenstates'' for
continuous spectrum observables (e.g., the rigged Hilbert space
formalism).  However, these constructions are not needed for my
purposes: Any continuous spectrum observable is associated with a
Boolean algebra of spectral projections, and a 2-valued homomorphism
on this Boolean algebra gives a generalized eigenstate.  Furthermore,
such homomorphisms should not be considered any less a part of the
Hilbert space formalism in the continuous case than they are in the
discrete case, where they just happen to be represented by proper
eigenvectors.

\subsection{Preliminaries}
I begin by establishing the terminology I will be using in this paper.
Let $\alg{B}$ be an arbitrary $\sigma$-complete Boolean algebra, with
additive identity $\bz$ and multiplicative identity $\mathbf{I}$.  We
say that $\omega$ is a \emph{state} of $\alg{B}$ just in case $\omega$
is a mapping of $\alg{B}$ into $[0,1]$ such that $\omega
(\mathbf{I})=1$ and \begin{equation} \omega (A\vee B)=\omega
  (A)+\omega (B) ,\end{equation} for each disjoint $A,B\in \alg{B}$,
i.e., $A\wedge B=\bz$.  Clearly, the set of states on $\alg{B}$ is a
convex set.  If $\omega$ is an extreme point among the states on
$\alg{B}$, we say that $\omega$ is a \emph{pure} state.  That is,
$\omega$ is pure just in case: If $\omega =a\rho +(1-a)\tau$, for some
$a\in (0,1)$ and states $\rho ,\tau$ on $\alg{B}$, then $\omega =\rho
=\tau$.  The state $\omega$ is pure if and only if it is a Boolean
homomorphism of $\alg{B}$ onto $\{ 0,1\}$.  If $\omega$ is not pure,
we say that $\omega$ is \emph{mixed}.  We say that a state $\omega$ is
\emph{normal} just in case
\begin{equation} \omega \left( \vee _{i=1}^{\infty}A_{i} \right) =\sum
  _{i=1}^{\infty}\omega (A_{i}) ,\end{equation} for each countable
pairwise disjoint sequence $\{ A_{i} \}$ in $\alg{B}$.  If $\omega$ is not
normal, we say that $\omega$ is \emph{singular}.  Recall that a family
$\alg{F}$ of elements in $\alg{B}$ is called a \emph{filter} just in
case: (i) $\emptyset \not\in \alg{F}$, (ii) $A\wedge B\in \alg{F}$,
when $A,B\in \alg{F}$, and (iii) if $A\in \alg{F}$ and $A\leq B$, then
$B\in \alg{F}$.  If, furthermore, either $A\in \alg{F}$ or $\neg A\in
\alg{F}$ for all $A\in \alg{B}$, then $\alg{F}$ is said to be an 
\emph{ultrafilter}.  Each ultrafilter $\alg{U}$ on a
Boolean algebra $\alg{B}$ is the preimage of $1$ under some pure state
$\omega$ of $\alg{B}$, and $\omega$ is normal if and only if $\alg{U}$
is closed under countable meets.  For any non-zero element $A\in \alg{B}$, the set 
$\alg{U}_{A}=\{ B\in \alg{B}:A\leq B \}$ is a filter, and is an ultrafilter if $A$ is an
atom in $\alg{B}$.  In this case, $\alg{U}_{A}$ is called the \emph{principal
ultrafilter} generated by $A$.  It is easy to see then that principal 
ultrafilters give rise to normal pure states.

The discussion of this paper focuses on two Boolean algebras: The
$\sigma$-complete Boolean algebra $\alg{B}(\mathbb{R})$ of Borel
subsets of $\mathbb{R}$, and its quotient $\quo$ by the ideal of
Lebesgue measure zero sets.  Recall that $\bor{\mathbb{R}}$ is the
(Boolean) $\sigma$-algebra generated by the open subsets of
$\mathbb{R}$.  The logical operations on $\bor{\mathbb{R}}$ are just
the set-theoretic operations $\cap$ (intersection), $\cup$ (union),
and $\sim$ (complement in $\mathbb{R}$).  Clearly, $\bor{\mathbb{R}}$
is purely atomic; i.e., for each $S\in \bor{\mathbb{R}}$ there is some
atomic element $A\in \bor{\mathbb{R}}$ (i.e., a singleton set) such
that $A\subseteq S$.  Thus, by the considerations adduced above,
$\bor{\mathbb{R}}$ has an abundance of pure normal states.

Let $\alg{N}$ denote the family of Lebesgue measure zero sets in
$\alg{B}(\mathbb{R})$.  We define an equivalence relation $\approx$ on
$\bor{\mathbb{R}}$ by setting $S_{1}\approx S_{2}$ just in case
$S_{1}\triangle S_{2}\in \alg{N}$, where $S_{1}\triangle
S_{2}=(S_{1}\backslash S_{2})\cup (S_{2}\backslash S_{1})$.  For each
$S\in \bor{\mathbb{R}}$, let $\pi (S)=\{ S'\in
\bor{\mathbb{R}}:S'\approx S\} $, and let
\begin{equation}
\quo = \left\{ \pi (S):S\in \bor{\mathbb{R}} \right\} . \end{equation}
For any countable family $\{ \pi (S_{i}) \}$ of elements in $\quo$, we
define, 
\begin{equation}
\vee _{i=1}^{\infty} \pi (S_{i}) =\pi \left( \cup _{i=1}^{\infty}
S_{i} \right) .\end{equation}  (This is well-defined since $\alg{N}$
is a $\sigma$-ideal.)  For $\pi (S)\in \quo$, we
set $\neg \pi (S)=\pi (\sim \!\!S)$.  It then follows that $\langle \quo 
,\vee ,\neg \rangle$ is a $\sigma$-complete Boolean algebra.  It is
also possible to show --- although I do not prove it here --- that
$\quo$ has \emph{no} atoms; i.e., for each nonzero $A\in \quo$, there is 
some nonzero $B\in \quo$ such that $B<A$.  This, in fact, is equivalent to the
nonexistence of pure normal states for $\quo$, which I do prove below
in Proposition~\ref{nosig}.  

\subsection{The position observable in QM}
It is well-known that no wavefunctions $\psi$ in a Hilbert space give
a probability density $\abs{\psi}^{2}$ that is focused at one point.
In this section, I will show that this well-known fact may be
described in purely logical terms as the nonexistence of pure normal
states on the Boolean algebra $\quo$.

The state space of quantum mechanics is obtained by requiring an
irreducible representation of the canonical commutation relation
(CCR): \begin{equation} \bigl[ \widehat{Q},\widehat{P} \bigr] =i\hbar
  I,\label{ccr}
\end{equation} by operators $\widehat{Q},\,\widehat{P}$ on some
Hilbert space $\hil{H}$.  In its integrated (Weyl) form, there is a
unique (up to unitary isomorphism) representation of the CCR on the
Hilbert space $L_{2}(\mathbb{R})$ of equivalence classes of square
integrable functions from $\mathbb{R}$ into
$\mathbb{C}$~\cite{summers,unique}.  (Two functions are equivalent
just in case they agree except possibly on a set of measure zero.)  In
this representation, the position operator $\widehat{Q}$ is defined on
a dense subset of $L_{2}(\mathbb{R})$ by the equation
\begin{equation}
(\widehat{Q} \psi )(q)= q\cdot \psi (q) \, .\end{equation}
A ``proposition'' about the position of the particle is then given by
a projection operator which is a function of $\widehat{Q}$.  In
particular, for each $S\in \bor{\mathbb{R}}$, we define an operator
$E(S)$ on $L_{2}(\mathbb{R})$ by the equation \begin{equation}
  (E(S)\psi )(q) = \chi _{S}(q) \cdot \psi (q) .\end{equation} The
set $\alg{O}_{Q}=\{ E(S):S\in \bor{\mathbb{R}} \}$ is a
$\sigma$-complete Boolean algebra of projection operators on the
Hilbert space $\hil{H}$, and $E$ is a Boolean
$\sigma$-homomorphism from $\bor{\mathbb{R}}$ onto $\alg{O}_{Q}$.
Moreover, $E(S)=\bz$ if and only if $S$ has Lebesgue measure zero.
Thus, the kernel of the mapping $E$ is $\alg{N}$, and (by the first
isomorphism theorem for rings) $\alg{O}_{Q}$ is isomorphic to the
quotient Boolean algebra $\bor{\mathbb{R}}/\alg{N}$.  ($E$ is usually 
called the spectral measure for $\widehat{Q}$.)  

Now, every wavefunction $\psi \in \hil{H}$ gives rise to a normal
state $\rho _{\psi}$ on $\alg{O}_{Q}$ in the standard way:
\begin{equation} \rho _{\psi}(E(S))\df \langle \psi |E(S)\psi
  \rangle =\int _{\mathbb{R}}\chi _{S}(q)\abs{\psi (q)}^{2}dq =\int
  _{S}\abs{\psi (q)}^{2}dq .\end{equation} However, it follows from
purely logical considerations that there cannot be normal
``eigenstates'' for $\widehat{Q}$.
 
\begin{prop} There are no pure normal states of $\quo$.  \label{nosig} \end{prop}
\begin{proof} Let $\omega$ be a pure state of $\quo$, and let
  $\alg{U}=\omega ^{-1}(1)$ be the corresponding ultrafilter.  (Case
  1) Suppose that $\pi (S)\not\in \alg{U}$ for all compact subsets $S$
  of $\mathbb{R}$.  For each $n\in \mathbb{N}$, let $A_{n}=[-n,n]$.
  Then, $\pi (A_{n})\not\in \alg{U}$ and since $\alg{U}$ is an
  ultrafilter, $\pi (\sim \!\!A_{n})=\neg \pi (A_{n})\in \alg{U}$.
  However,
  \begin{equation} \wedge _{n=1}^{\infty} \pi (\sim \!\! A_{n}) = 
\pi \left( \cap _{n=1}^{\infty} \,\sim \!\! A_{n} \right) =\bz
,\end{equation} and therefore $\alg{U}$ is not closed under countable
meets.  (Case 2) If $\pi (S)\in \alg{U}$ for some compact set $S$ in
$\mathbb{R}$, then $\alg{U}$ contains $\pi (\alg{F}_{\lambda })$ where
$\alg{F}_{\lambda }$ is the family of open neighborhoods of some point $\lambda
\in \mathbb{R}$.  For each $n\in \mathbb{N}$, let \begin{equation}
  B_{n}=\left( \lambda -n^{-1},\lambda +n^{-1} \right) \, .
\end{equation} Then, $\pi (B_{n})\in \alg{U}$ for each $n\in
\mathbb{N}$.  However, 
\begin{equation} 
\wedge _{n=1}^{\infty} \pi
  (B_{n}) =\pi \left( \cap _{n=1}^{\infty}B_{n}\right) =\pi
  \left( \{ \lambda \} \right) =\bz ,\end{equation} since $\{
\lambda \}$ has Lebesgue measure zero.  Thus, $\wedge _{n=1}^{\infty}
\pi (B_{n})\not\in \mathcal{U}$ and $\mathcal{U}$ is again not closed
under countable meets.
\end{proof}

Thus, assuming that quantum theory is descriptively complete, and that
quantum states must be $\sigma$-additive, $\abs{\psi}^{2}$ can not be
interpreted as a measure of our ignorance of the true (pure) state of
individual particles in the ensemble.  Some philosophers draw the
moral from this that quantum theory is teaching us something new about
the nature of continuous quantities, namely that we should not think
or talk about continuous quantities possessing point values, nor
should we think or talk about statistical states as distributions over
point-valued states.  For example, Fine~\cite{fine} argues that we
should think of statistical state $\rho _{\psi}$ as a map that assigns a
\emph{set}, rather than a numerical value, to $\widehat{Q}$.  In
particular,
\begin{equation} \rho _{\psi}(\widehat{Q}) \df  \bigcap
  \left\{ \,S\,:\, \rho _{\psi} (E(S))=1 \,\right\}
  .\end{equation}
Of course, $\rho _{\psi}(\widehat{Q})$ will never be a singleton,
  and --- perhaps contrary to our intuitions --- the statement that
  the particle lies in $S$ is a maximally specific description of the location of the particle.  
  
  Teller~\cite{teller} agrees with Fine that a statistical state
  should be interpreted not as a mixture of value states, but as
  assigning a ``partless'' set value.  Nonetheless, Teller does
  recognize that this description of the location of the particle can
  be misleading: it suggests --- contrary to the predictions of
  quantum theory --- that no matter where we look in $S$, we will find
  a ``piece'' of the particle.  Teller suggests, however, that we
  should think of the property ``being located in $S$'' as including,
  \begin{quote} ``a collection of dispositions or potentialities to
    manifest a more refined position of the same nature whenever a
    more refined measurement interaction takes place.'' (p. 357)
  \end{quote} Thus, Teller seems to be claiming that the probability
  distribution $\abs{\psi}^{2}$ itself gives a maximally specific
  description of an individual element in the ensemble.
  
  Teller claims, moreover, that this dual property/disposition nature
  of position in QM is ``nothing exceptional;'' in fact, it is just
  like the color white which is ``a manifest property and at the same
  time an array of dispositions''~(ibid).  This analogy to color,
  however, seems to show exactly what leaves us uneasy about Teller's
  proposal.  In the case of the color white, we can explain an
  object's dispositions to appear certain ways in certain contexts in
  terms of categorical properties of the object (viz., the chemical
  composition of its surface).  On the other hand, Teller is claiming
  that a quantum-mechanical particle can have irreducibly
  probabilistic dispositions with no categorical basis (apart from,
  perhaps, $\psi$ itself, which is difficult to see as anything other
  than a catalog of those dispositions).

\subsection{The ignorance interpretation in classical mechanics}
According to Teller, these irreducibily probabilistic dispositions are
a peculiar feature of quantum mechanics.  Unlike the quantum case, the
probabilities in classical mechanics may be thought of as measures of
our ignorance of the precise state:
\begin{quote} ``In the context of classical physics the outcome of an
  inexact measurement may be described as a probability distribution
  over exact values, and we know how to state what this means: There
  is a probability (interpreted as a relative frequency, subjective
  degree of belief, or propensity) for the value of the measured
  quantity to have one or another of the exact values in the support
  of the distribution.'' (p. 352) \end{quote}
Let us attempt, then, to give a mathematically rigorous description of
this purported distinction between the classical and quantum cases.

A classical particle moving in one dimension has a \emph{phase space}
$\mathbb{R}^{2}$, with one coordinate for the position $Q$ of the
particle, and the other coordinate for the momentum $P=m\cdot (dQ/dt)$
of the particle.  We may think of Borel subsets of $\mathbb{R}^{2}$ as
representing statements that ascribe properties to the particle.  (For
background material, see Chap. 1 of Ref.~\ref{vara}.)  A dynamical
variable is represented by a measurable, real-valued function $F$ on
phase space.  For example, the position of the particle is represented
by the function $F((q,p))=q$.  For any Borel set $S$ in $\mathbb{R}$,
$F^{-1}(S)$ consists precisely of those ``states'' (i.e., points in
phase space) for which the value of $F$ lies in $S$.  $F^{-1}$
(considered as a set mapping) is in fact a Boolean
$\sigma$-homomorphism from $\bor{\mathbb{R}}$ onto some Boolean
subalgebra $\alg{B}_{F}$ of $\bor{\mathbb{R}^{2}}$.  For example, the
disjunctive proposition \emph{``The value of $F$ lies either in
  $S_{1}$ or in $S_{2}$''} is represented formally by
\begin{equation} F^{-1}(S_{1}\cup S_{2})=F^{-1}(S_{1})\cup F^{-1}(S_{2}).  \end{equation}   

Consider now the completely classical example of a person (or machine)
Bob throwing darts at a (one-dimensional) dartboard.  Since we are
ignorant of the factors that influence each individual throw, we
assign a probability density $\rho$ to the points on the dartboard
(represented here by $\mathbb{R}$).  Now, just as in the
quantum-mechanical case, the distribution $\rho$ on $\mathbb{R}$
defines a mixed normal state on $\alg{B}_{F}$ by means of the equation
\begin{equation} \rho (F^{-1}(S))=\int _{\mathbb{R}}\chi _{S}(q) \rho
  (q) \,dq =\int _{S}\rho (q) \, dq.\end{equation} 
Any particular dart in the ``ensemble'' should, however, be characterized by a
\emph{pure} state on $\alg{B}_{F}$.  (i.e., there is some point
$\lambda \in \mathbb{R}$ such that the dart has probability $1$ of
hitting that point).  And, in fact, everything works out fine in this
case since there \emph{are} pure normal states of $\alg{B}_{F}$.
Indeed, the atoms in
$\alg{B}_{F}$ are just lines in $\mathbb{R}^{2}$ on which $F$ is constant:
$A_{\lambda}=\{ (q,p):F((q,p))=\lambda \}$.  Since each atom in a
Boolean algebra gives rise to a pure normal state, there is a normal
pure state $\omega _{\lambda }$ of $\alg{B}_{F}$ defined explicitly by
the condition that $\omega _{\lambda}(S)=1$ if and only if
$A_{\lambda}\subseteq S$.  Conversely, if $\omega $ is a pure normal
state of $\alg{B}_{F}$, then $\omega =\omega _{\lambda }$ for some
$\lambda \in \mathbb{R}$.  [First show that $\omega (F^{-1}(S))=1$ for
some compact set $S$.  Standard topological arguments then show that
$\omega (F^{-1}(V))=1$ for every open neighborhood $V$ or some point 
$\lambda \in \mathbb{R}$.  Now use the fact that $\{ \lambda \}$ is 
the intersection of a countable family of open sets.]  Thus, the pure 
normal states on $\alg{B}_{F}$ are in one-to-one correspondence with 
the sets on which $F$ takes a constant value.  Finally, it is easy to
see that the distribution $\rho$ does admit interpretation as a measure of ignorance
over the pure normal states, namely $\rho =\int
\omega _{\lambda }d\rho $.  

However, before we grant that there is a fundamental difference here
between the quantum and classical cases, let's think a bit more
carefully about the interpretation of the mathematical objects we've
been dealing with.  In particular, an element $F^{-1}(S)\in
\alg{B}_{F}$ is interpreted as a statement that \emph{ascribes a
  property} to the particle, namely the statement \emph{``The particle
  is located in $S$.''}  Now take an element $E(S)\in \alg{O}_{Q}$.
Is it the quantum equivalent of the classical statement $F^{-1}(S)$;
i.e., an ascription of a property \emph{``located in S''}?  If we
interpret $E(S)$ this way, then it is trivially true that a quantum
particle cannot have a sharp location:
\begin{quote}
  Let $\lambda \in \mathbb{R}$.  Then $E(\{ \lambda \})=\bz$ since $\{
  \lambda \}$ has Lebesgue measure zero.  Thus, the proposition
  \emph{``The particle is located at $\lambda$''} is a contradiction.
  \emph{QED} \end{quote} Thus the property ascription interpretation
of elements of $\alg{O}_{Q}$ rules out our being able to describe
particles with sharp locations.  However, the property ascription
interpretation is not the traditional interpretation of projection
operators in quantum theory.  According to the traditional
interpretation, the projection operators are \emph{observables} or
\emph{experimental propositions}.  This is sometimes made precise by
saying that $E(S)$ represents the statement:
\begin{quote} \emph{``A measurement of $\widehat{Q}$ will yield a value lying in $S$.''}
\end{quote}
Under this interpretation, $E( \{ \lambda \} )=\bz$ does \emph{not}
entail that a particle cannot be located at $\lambda$; it simply means
that it is false (in fact, a contradiction) to say that any
measurement of $\widehat{Q}$ will yield the precise value $\lambda$.
(This, for example, may be the result of the fact that precise
measurements of $\widehat{Q}$ are impossible.)  But now, if
$\alg{O}_{Q}$ consists of observables, while $\alg{B}_{F}$ consists of
property ascriptions, differences in the state spaces of the two
logics do not necessarily reflect differences in the descriptive
resources of classical and quantum mechanics.

The distinction between observables (experimental propositions) and
property ascriptions should not be drawn only in quantum physics.  For
example, although $F^{-1}(\lambda )\in \alg{B}_{F}$ is a classical
property ascription, it is doubtful that any classical experiment ever
determines that the value of the quantity $F$ is the \emph{exact} real
number $\lambda$; and thus the proposition ``$\val{F}=\lambda$'' is
not really an experimental proposition --- even in the classical case.
The same intuition is expressed by Birkhoff and von
Neumann~\cite{birkhoff} who claim that, it would be absurd,
\begin{quote} ``to call an ``experimental proposition,'' the assertion
  that the angular momentum (in radians per second) of the earth
  around the sun was at a particular instant a rational
  number!''~(p.~825) \end{quote} 
Indeed, what experiment would we perform in order to determine that it
  was a rational number?  

There will certainly be a number of different ways to give a
mathematically precise account of the notion of approximate
measurement.  Here, though, I will just explain briefly von
Neumann's~\cite[595-598]{operate} account, without trying to argue
positively for its merits.  I wish simply to show that according to
von Neumann's account of approximate measurement in \emph{classical}
physics, the logic of classical experimental propositions (for one
continuous quantity) is mathematically identical to $\alg{O}_{Q}$.

The fact that measurements are of arbitrarily good, but always finite,
precision is given formal statement by von Neumann as follows:

\begin{quote} For each $n\in \mathbb{N}$, the value space $\Omega$ of $F$ may
  be partitioned into a finite or countably infinite family of
  measurable sets $\{ N_{i}^{(n)} :i=1,2,\dots \}$, of which each two
  have measure-zero overlap.  We may do this so that the partition $\{
  N_{i}^{(n+1)} \}$ is strictly finer than $\{ N_{i}^{(n)} \}$ for
  each $n$, and if we let $\delta _{n}$ denote the maximum diameter of
  the sets in $\{ N_{i}^{(n)} \}$, then $\delta _{n}\rightarrow 0$.
  We say that some procedure is a $F$-measurement of $n$-th level
  precision just in case it determines the $i\in \mathbb{N}$ such that
  $\val{F}\in N_{i}^{(n)}$.  Then, $F$-measurements of all finite
  levels of precision are possible.  \end{quote}
  
With this as the basic principle of measurement, von Neumann proves:
For any set $S\in \mathbb{R}$, we can determine via measurement (up to
an arbitrary level of confidence) whether $\val{F}\in S$, if and only
if $S$ has non-vanishing Lebesgue measure.  Accordingly, it is not
possible to determine with any good level of confidence that $\val{F}$
lies in a Lebesgue measure zero set --- in agreement with the
intuition that ``$\val{F}=\lambda$'' and \emph{``$\val{F}$ is a
  rational number''} are not experimental propositions.  Furthermore,
if two subsets $S_{1},\, S_{2}\in \alg{B}(\mathbb{R})$ agree except
possibly on a set of measure zero, then there is no way to determine
that $\val{F}\in S_{1}\backslash S_{2}$ or that $\val{F}\in
S_{2}\backslash S_{1}$.  Thus, two elements $S_{1},S_{2}\in
\alg{B}_{F}$ define the same classical observable just in case
$S_{1}\triangle S_{2}=(S_{1}\backslash S_{2})\cup (S_{2}\backslash
S_{1})\in \alg{N}$.  But then the logic of classical observables
$\alg{O}_{F}$ for the quantity $F$ is \emph{also} the quotient Boolean
algebra $\quo$. (See p.~825 of Ref.~\ref{birkhoff}.)

Now, Teller claims that,
\begin{quote} 
  ``...if we believe that systems possess exact values for continuous
  quantities, classical theory contains the descriptive resources for
  attributing such values to the system, whether or not measurements
  are taken to be imprecise in some sense.  Quantum mechanics has no
  such descriptive resource''~(p.~352).
\end{quote}
Apparently, then, Teller endorses the claim that classical mechanics
does have states assigning exact values to a continuous quantity, even
after we take into account the fact that absolutely precise
measurements are not possible.  However, when we take that into
account, the logic of classical experimental propositions
$\alg{O}_{F}$ is \emph{isomorphic} to the logic of quantum
experimental propositions $\alg{O}_{Q}$.  Moreover, if we require
quantum states to be $\sigma$-additive (thereby depriving quantum
theory of the resources to describe exact values), we should also
require classical states to be $\sigma$-additive --- which would also
deprive classical theory of the resources to describe exact values.
But this conclusion cannot be right: Classical theory does already
have these resources to describe exact values, namely points in phase
space.  The difficulty, then, is not with a lack of resources; but
rather with the connection between the theoretical description of
point values (viz., pure normal states on $\alg{B}_{F}$) and the
``surface'' description given by states on $\alg{O}_{F}$.  I will
argue in the next section that the solution to this difficulty is that
the ``hidden states'' give rise to finitely-additive (singular) states
on $\alg{O}_{F}$.

\subsection{Singular states}
Let $\alg{S}$ be any family of elements in $\quo$ with the finite meet
property.  That is, for any finite family $\{ A_{1},\dots ,A_{n}
\}\subseteq \alg{S}$, we have \begin{equation} A_{1}\wedge \cdots
  \wedge A_{n}\neq \bz .\end{equation} Then, the Ultrafilter Extension
Theorem~\cite[p.~339]{schec} entails that there is some 2-valued
homomorphism $\omega$ on $\quo$ such that $\omega (A)=1$ for all $A\in
\alg{S}$.  In particular, $\quo$ \emph{does} have pure states, even
though they are not $\sigma$-additive.  Furthermore, any probability
distribution $\rho$ on $\quo$ can be written explicitly as a mixture
of pure states; i.e., there is some measure $\mu$ on the pure state
space of $\quo$ such that
\begin{equation}
\rho =\int \omega _{\alpha}\,d\mu (\alpha ) \,. \label{enough} \end{equation}  
Thus, the distribution $\rho$ does admit an ignorance interpretation;
viz., it measures our ignorance of the (singular) state which
describes an individual particle in the ensemble.

The pure states on $\quo$ should be thought of as ``surface states.''
That is, each pure state on $\quo$ gives a consistent catalog of
sure-fire responses of the system to all possible measurements.  I will
now argue that for each $\lambda \in \mathbb{R}$, there is at least
one surface state whose predictions can be interpreted as consistent
with the fact that $\val{F}=\lambda$, and inconsistent with the fact
that $\val{F}=\xi$ for any $\xi \neq \lambda$.

Consider again the dart throwing example.  According to the von
Neumann measurement theory, there is no experiment which would tell us
the real number $\lambda$ at which the dart lands.  However, we can
rule out segments of the dart board.  In particular, choose some open
segment $V$ along the line.  Then, if the system responds $0$ to the
measurement $\pi (V)$, we are entitled to conclude that the dart did
not land at any point in the segment $V$.  In other words, a necessary
condition for the value of $F$ to be $\lambda$ in the state $\omega$
is that:
\begin{eqnarray}
\omega (\pi (V))=1,\:\mbox{for all} \; V\in \alg{F}_{\lambda}
\nonumber \qquad \qquad \qquad (*) \end{eqnarray}
where $\alg{F}_{\lambda}$ is the family of open neighborhoods of $\lambda$.
Suppose, conversely, that $\omega$ satisfies $(*)$ with respect to
some point $\lambda \in \mathbb{R}$.  [If $(*)$ holds, we say that the
state $\omega$ converges to the point $\lambda$.]  Then, for any other
point $\xi$ along the line, there is some open neighborhood $W$ of
$\xi$ which excludes $\lambda$.  A measurement of $\pi (W)$ in
the state $\omega$ then shows that the dart did not land at $\xi$.  Thus, when
$\omega$ converges to $\lambda$, the only interpretation consistent
with the von Neumann theory of measurement is that $\val{F}=\lambda$;
that is, the ``hidden'' state of the system is given by some phase
space point along the line $\{ (q,p):F((q,p))= \lambda \}$.

Now consider again some mixed state $\rho$ on $\quo$.  Then, $\rho$ is
a mixture of pure states as in Eq.~\ref{enough}, and each pure state
assigns some definite value to $F$.  (We do have to be sure that
$\rho$ falls off sufficiently quickly at infinity.  See Sec.~3 of
Ref.~\ref{fink}.)  Thus, this validates the idea that in classical
mechanics, the distribution $\rho$ represents the ``...probability for
the value of the measured quantity to have one or another of the exact
values in the support of the distribution.''

However, everything we have just said about singular states of $\quo$
was entirely neutral as to whether we were discussing classical or
quantum mechanics.  Thus, a quantum-mechanical probability
distribution $\abs{\psi}^{2}$ may also be thought of as representing a
probability for $\widehat{Q}$ to have one or another of the exact
values in the support of $\abs{\psi}^{2}$.  According to Teller,
though, even if there were some mathematically acceptable method for
describing sharp positions in QM, doing so would not be
\emph{physically} acceptable:
\begin{quote} ``Such an extended physical theory would describe
  systems having totally indeterminate momentum, not even highly
  localized to any finite interval.  Such systems would have an
  infinite expectation value for their kinetic energy''~(p.~353).
\end{quote} 
This argument, however, is not a criticism of singular states
\emph{per se} (which, by the way, are not extensions of the standard
formalism).  Since the energy observable is an unbounded operator,
there are vectors in the Hilbert space that are not in the domain of
this observable, and so these vectors too will assign ``infinite
kinetic energy'' to the system.  Thus, following Teller's reasoning
here to its logical conclusion, the Hilbert space is not really the
state space, rather the domain of the energy observable is the state
space.  But then, what about other unbounded observables such as the
position observable $\widehat{Q}$ itself?  Should we also throw away
all those vectors that assign an infinite expectation value to
position?  And once we've cleansed the Hilbert space of vectors not in
the domains of all physically relevant observables, would anything be
left?

Besides these obvious retorts, it is not completely clear what is
meant by Teller's talk about assigning infinite expectation values.
What we do know --- in the case of pure (singular) states of
$\alg{O}_{Q}$ --- is that these states cannot be thought of as
assigning some finite expectation values to the energy observable.
(Precisely: Each pure, convergent state on $\alg{O}_{Q}$ gives rise to
a mixed state on the logic of experimental propositions about energy,
and measure one of the pure states in its integral decomposition do
not converge to a finite value.  See Ref.~\ref{fink}, Section~3.)  It
may be mathematically convenient, then, to adjoin $\infty$ to the
normal range of values, but we should \emph{not} say that these states
assign ``$\infty$'' to the energy observable.  Rather, we should think
of this infinity --- as we do of the infinities related to
singularities in General Relativity --- as indicating the limits of
the descriptive capabilities of our theory.  And if, with Teller, we
take the theory to be descriptively complete, then there is nothing
more to describe about energy in these situations.

\section{Potential difficulties with singular states}
In the previous section, I have argued that singular states permit us
to give an ignorance interpretation of the probability distributions
associated with individual continuous quantities in QM.  I imagine my
reader may still not be convinced, though, of the acceptability of
singular states.  In this section, I will confirm that there are
indeed difficulties with these singular states.  In the final section,
however, I will show that these difficulties can be remedied by taking
a more modest view of the measurements that can be made on a
continuous quantity.

It would be quite natural to think that if we pick a point $\lambda
\in \mathbb{R}$, it determines uniquely a pure state $\omega$ on
$\quo$ that converges to $\lambda$.  There is, however, a difficulty.
Suppose we divide the screen (or dartboard) into two halves $A=\{
q:q>\lambda \}$ and $B=\{ q :q<\lambda \}$.  Then,
\begin{equation}
\pi (A)\vee \pi(B) =\pi (A)\vee \pi (B) \vee \pi (\{ \lambda \} )=\pi
(\mathbb{R})=\mathbf{I} ,\end{equation}
since $\{ \lambda \}$ has Lebesgue measure zero.  Thus, \begin{equation}
\omega (\pi (A))+\omega (\pi (B)) =1 
,\end{equation}
while either $\omega (\pi (A))=0$ or $\omega (\pi (B))=0$.  Thus, any
pure state $\omega$ must assign $1$ to either $\pi (A)$ or $\pi (B)$,
but not to both.  What would be the right answer for a state in which
the particle is located at $\lambda$?      

Actually, it is easy to show that both value assignments are
consistent with the particle being located at $\lambda$.  Indeed, let
$\alg{S}_{1}=\alg{F}_{\lambda}\cup \{ \pi (A) \}$ and let
$\alg{S}_{2}=\alg{F}_{\lambda }\cup \{ \pi (B) \}$.  Since both
$\alg{S}_{1}$ and $\alg{S}_{2}$ have the finite meet property, there
is an ultrafilter $\alg{U}_{1}$ which contains $\alg{S}_{1}$ and an
ultrafilter $\alg{U}_{2}$ which contains $\alg{S}_{2}$.  Since
$\alg{U}_{1}$ and $\alg{U}_{2}$ both converge to $\lambda$, both
should be interpreted as surface manifestations of some ``hidden''
state in which the particle is located at $\lambda$.  On the other
hand, the particle's being in some hidden state $(\lambda ,p)$, gives
no information concerning its disposition to respond to a measurement
of $\pi (A)$.

The apparent mismatch between hidden states and surface states is, in
fact, quite severe.  

\begin{prop} For each $\lambda \in \mathbb{R}$, there are at least
  $\aleph _{0}$ distinct ultrafilters on $\quo$ that converge to
  $\lambda$. \label{many} \end{prop}

In order to prove this proposition, we will first require a lemma.

\begin{lemma}  Suppose that $\{ B_{j}:j\in J \}$ is a
  family of open subsets of $\mathbb{R}$ such that $B_{i}\cap
  B_{j}=\emptyset$ when $i\neq j$ and such that $\lambda \in
  \clo{B_{j}}$ for all $j\in J$.  Then, there is a family $\{
  \mathcal{U}_{j}:j\in J \}$ of distinct ultrafilters on $\quo$ such
  that $\pi (\mathcal{F}_{\lambda })\subseteq \mathcal{U}_{j}$ for all
  $j\in J$. \label{disjoint} \end{lemma}

\begin{proof} Let $\{ B_{j}:j\in J\}$ be given as above.  For each $j\in J$,
  let \begin{equation} \mathcal{S}_{j}=\pi (\mathcal{F}_{\lambda
      })\cup \{ \pi (B_{j}) \} .\end{equation} From the Ultrafilter
  Extension Theorem, $\mathcal{S}_{j}$ is contained in a Boolean
  ultrafilter $\mathcal{U}_{j}$ if and only if the meet of any finite
  collection of elements in $\mathcal{S}_{j}$ is nonzero.  Let $\{ \pi
  (U_{1}),\pi (U_{2}),\dots ,\pi (U_{n}) \} \subseteq
  \mathcal{S}_{j}$.  If $U_{i} \in \mathcal{F}_{\lambda }$ for each
  $i$, then $\cap _{i=1}^{n}U_{i}$ is an open set that contains
  $\lambda $.  If $U_{1}=B_{j}$, and $U_{i}\in \mathcal{F}_{\lambda }$
  for $i\geq 2$, then $\cap _{i=2}^{n}U_{i}$ is an open set that
  contains $\lambda $, and since $\lambda \in \clo{B_{j}}$, $\cap
  _{i=1}^{n}U_{i}$ is a nonempty open set.  In either case, the open
  set $O\df \cap _{i=1}^{n}U_{i}$ is nonempty.  Since Lebesgue measure
  does not vanish on any nonempty open set, we have
  \begin{equation} \wedge _{i=1}^{n}\pi (U_{i}) =\pi \left( \cap
      _{i=1}^{n}U_{i} \right) = \pi (O)\neq 0.\end{equation}
  Thus, there is an ultrafilter $\mathcal{U}_{j}$ that contains
  $\mathcal{S}_{j}$.  Since $\pi (B_{i})\wedge \pi (B_{j})=0$ when
  $i\neq j$, it follows that $\mathcal{U}_{i}\neq \mathcal{U}_{j}$
  when $i\neq j$.  \end{proof}

\begin{proof}[Proof of the proposition] We will consider the
  case where $\lambda =0$.  In light of the lemma, it will suffice to
  construct a family $\{ B_{n}:n\in \mathbb{N} \}$ of open subsets of
  $\mathbb{R}$ such that $B_{n}\cap B_{m}=\emptyset$ when $n\neq m$
  and such that $0\in \clo{B_{n}}$ for all $n\in \mathbb{N}$.
  
  For each $n\in \mathbb{N}\cup \{ 0\}$, let $a_{n}=2^{-n}$.  For each
  $n,m\in \mathbb{N}$, let $b_{nm}=(a_{n}+a_{n-1})/2^{m}$, and let
\begin{equation}    A_{m}=\bigcup _{n=1}^{\infty} (a_{n},b_{nm}) .\end{equation} 
Note that $A_{m}\supset A_{m+1}$ for all $m\in \mathbb{N}$, and in fact
$A_{m}-A_{m+1}$ has nonempty interior.  Thus, if we set 
  \begin{equation}
B_{m}=\inte{A_{m}-A_{m+1}} ,\end{equation}  
then $B_{m}$ is a nonempty open set.  We must show that $B_{i}\cap
B_{j}=\emptyset$ when $i\neq j$ and that $0\in \clo{B_{i}}$ for all
$i$.  Suppose that $i\neq j$.  We may assume that $i+1\leq j$, in which case
$A_{i+1}\supseteq A_{j}$, 
and \begin{equation}
B_{i}\cap B_{j}\subseteq \left( A_{i}-A_{i+1}\right) \cap \left( 
A_{j}-A_{j+1}\right) \subseteq A_{j}-A_{i+1}=\emptyset
.\end{equation}
It is easy to verify that the sequence $\{ b_{nm}:n\in \mathbb{N} \}$ is in
$\clo{B_{m}}$ for each $m$.  But $\lim _{n}b_{nm}=0$ and since
$\clo{\clo{B_{m}}}=\clo{B_{m}}$, it follows that $0\in \clo{B_{m}}$.   
\end{proof}

Thus, even though we are able to explain probabilistic dispositions as
measures of ignorance of sure-fire dispositions, we are not able to
further reduce the latter to categorical location properties.  So, one
could still think that attributing categorical location properties is
objectionable on the grounds that these attributions are explanatorily
bankrupt.  

There is also a serious practical obstacle for using singular states
as the explanans for a reduction of probabilistic dispositions;
namely, none of these singular states can be explicitly defined.  By
saying that these singular states cannot be ``explicitly defined,'' I
mean to contrast this with the pure normal states of
$\bor{\mathbb{R}}$ given by principal ultrafilters, as well as with
mixed normal states on $\quo$.  First of all, it is quite trivial to
give an explicit definition of a principal ultrafilter on
$\bor{\mathbb{R}}$.  Indeed, once I choose some $\lambda \in
\mathbb{R}$, then (trivially) if you give me a Borel set $S$, then I
can tell you whether or not $S$ is in the principal ultrafilter
generated by $\{ \lambda \}$.  We have also seen that any
quantum-mechanical wavefunction $\psi$ defines a mixed normal state
$\rho _{\psi}$ on $\quo$ by means of the formula
\begin{equation} \rho _{\psi}(\pi (S))=\int _{S}\abs{\psi (q)}^{2}dq.
  \label{wind} \end{equation} Since it is possible to explicitly
define a wavefunction $\psi$, Eq.~\ref{wind} gives an explicit recipe
for computing the value $\rho _{\psi}(\pi (S))$ for any Borel set $S$.

On the other hand, although we ``know'' that there are ultrafilters
(i.e., pure states) on $\quo$, we do not know this because someone has
constructed an example of such an ultrafilter.  In order to obtain an
ultrafilter, we note that a certain (explicitly defined) family
$\mathcal{S}$ of elements in $\quo$ has the finite meet property, and
then we invoke the Ultrafilter Extension Theorem to extend $\alg{S}$
to an ultrafilter $\alg{U}$.  This extension procedure is a classic
example of nonconstructive mathematics: We are told that there is some
pure state $\omega$ on $\quo$, but we are not given a recipe for
determining the value $\omega (A)$ for an arbitrary element $A\in
\quo$.

A strong case can be made that the sometimes imprecise distinction
between ``giving an explicit example'' and ``proving existence
(nonconstructively),'' corresponds to a precise distinction in the
strength of the set-theoretic axioms used to prove the existence of
the object in question (see Ref.~\ref{schec}).  In particular, many
concrete mathematical objects (such as the Real
numbers~\cite[Appendix]{yanni}) are constructed using the Principle of
Recursive Constructions:
\begin{quote} (PRC) {\it Suppose $X$ is a non-empty set, and let a
    function $G$ be given from the set of finite sequences in $X$ into
    $\mathcal{P}(X)\backslash \{ \emptyset \} $.  Then there exists exactly one
    function $F:\mathbb{N}\mapsto X$ such that $F(0)\in G(\emptyset )$
    and $F(n)\in G(F(0),\dots ,F(n-1))$ for all $n>0$.} \end{quote}
PRC is equivalent in ZF to the axiom of Dependent
Choices~\cite[p.~147]{just}:
\begin{quote} (DC) {\it If $R$ is a binary relation on a nonempty set $X$ such
  that for every $x\in X$ there exists a $y\in X$ with $\langle
  x,y\rangle \in R$, then there exists a sequence $(x_{n})_{n\in
    \mathbb{N}}$ of elements of $X$ such that $\langle
  x_{n},x_{n+1}\rangle \in R$ for every $n\in \mathbb{N}$.}
  \end{quote}
  DC is a very weak form of the Axiom of Choice: ZF+DC entails the
  Axiom of Countable Choice, but ZF+DC does not entail the Ultrafilter
  Extension Theorem, nor does it entail any of the ``paradoxical''
  consequences of ZF+AC (e.g. the Banach-Tarski
  paradox)~\cite[Chap.~6]{schec}.  In fact, a good case can be made
  that \emph{applied} mathematics makes use \emph{only} of those
  objects in the ZF+DC universe~\cite{wright}.  For example, all
  theorems of classical (19th century) analysis, and even all theorems
  of the traditional Hilbert space formalism of quantum mechanics can
  be proved in ZF+DC~\cite{garnir}.
  
  Following~\cite{schec}, let us say that a mathematical object is
  \emph{intangible} just in case that object exists in the ZF+AC
  universe, but that object cannot be proved to exist in the ZF+DC
  universe.  For example, a free ultrafilter on $\mathbb{N}$ (i.e., an
  ultrafilter containing all subsets of $\mathbb{N}$ whose complements
  are finite) is an intangible~\cite{pincus}.  We now show that value
  states of $\quo$ are also intangibles.

\begin{prop}  It cannot be proved in ZF+DC that there is a pure state
  of $\quo$.  \label{nogo} \end{prop}

\begin{proof}  Let WUF (weak-ultrafilter principle) denote the
  statement that there is a free ultrafilter on $\mathbb{N}$, and let
  PS denote the statement that there is a pure state on $\quo$.  We
  must show that ZF+DC+$\neg$PS is consistent.  Since ZF+DC+$\neg $WUF
  is consistent~\cite{pincus}, it will suffice to show that
  ZF+DC$\,\vdash\,$(PS$\,\rightarrow\,$WUF).  Thus, suppose that PS
  holds; i.e., there is an ultrafilter $\mathcal{U}$ on $\quo$.  By
  Prop.~\ref{nosig}, $\mathcal{U}$ is not closed under countable
  meets; i.e., there is a family $\{ P_{n}:n\in \mathbb{N} \}$ of
  elements of $\quo$ such that $P_{n}\in \mathcal{U}$ for all $n\in
  \mathbb{N}$, but $\wedge _{n\in \mathbb{N}}P_{n} \not\in
  \mathcal{U}$.  (It's important to note that Prop.~\ref{nosig} uses
  no choice axiom stronger than DC.)  Define a family $\mathcal{J}$ of
  subsets of $\mathbb{N}$ by
  \begin{equation} J\in \mathcal{J} \quad \Longleftrightarrow \quad \wedge
  _{n\in J}P_{n}\not\in \mathcal{U}.\end{equation} It is easily verified
  that $\mathcal{J}$ is an ultrafilter in $\mathcal{P}(\mathbb{N})$ that
  contains all subsets of $\mathbb{N}$ whose complements are finite.
  \end{proof}

\section{The logic of unsharp experimental \\ propositions}
I have argued that it is possible to give an ignorance interpretation
of the probabilites associated with a continuous quantity in quantum
mechanics.  In order to do so, however, I had to make use of singular
states, which turn out to have some pretty odd properties that
militate against their playing a serious role in explaining a
particle's probabilistic dispositions.  Some might attribute the
oddities encountered in the last section to the fact that we permitted
ourselves to use states that are not $\sigma$-additive.  On the
contrary, I will now argue that the difficulty is not with the states
but with the observables: We were trying to account for too many
measurement outcomes.

If we take $\quo$ to be in one-to-one correspondence with possible
experiments, then Prop.~\ref{many} shows that there are experiments
whose outcomes could distinguish two states, both of which correspond
to one point $\lambda \in \mathbb{R}$.  If, however, these
distinguishing experiments cannot in fact be performed, then these two
states corresponding to $\lambda$ are empirically equivalent.  In this
section, I will show how to make this idea precise using the fact that
real experimental questions are always ``unsharp.''

Consider again a person, Bob, throwing darts at a one-dimensional
dartboard.  With the knowledge we have of Bob's skills, we choose a
region $S$ on the dartboard which we feel confident Bob can hit on 100
consecutive throws.  Since we don't have time, though, to watch each
of Bob's throws, we rig a device to keep score.  In particular, each
time a dart hits the board, we get a printout that records $y$ when
the dart lands in $S$ and that records $n$ when the dart lands outside
of $S$.  Now, after Bob has made his $100$ throws, we take a look at
the printout: there are $98$ $y$'s and $2$ $n$'s.  Do we then conclude
that Bob was not as skilled a dart thrower as we thought?  Not
necessarily.  Perhaps it was our score-keeping device, not Bob, which
failed on those two counts.  Or, perhaps some environmental factor
interfered with two of Bob's throws.  In any case, the point is that
in a real experimental situation, the measuring device itself and the
environment introduce factors of uncertainty which have to be taken
into account when interpreting out measurement results.

To make the imperfections in the experimental setup mathematically
precise within the confines of the standard Hilbert space formalism of
QM, we can associate a confidence measure $e$ with the device
$\alg{M}$.  (Here I follow Busch et al.~\cite{busch}.)  If, using this
measuring device, we attempt to test for the claim ``the particle is
in $S$,'' then what we actually measure is the ``smeared'' observable
\begin{equation}
E^{e}(S)\df  \int _{\mathbb{R}} e(q)\, E(S+q)\, dq =(\chi
_{S}*e)(\widehat{Q}).\end{equation}
Here the operation $*$ is the convolution product defined by 
\begin{equation}
(f*g)(q_{0})\df  \int _{\mathbb{R}}f(q)\,g (q_{0}-q) \,dq . 
\end{equation}
For a fixed confidence measure $e$, the mapping $S\mapsto E^{e}(S)$ is
a positive operator valued (POV) measure on $(\mathbb{R},\bor{\mathbb{R}} )$.
If $S_{1}$ and $S_{2}$ are disjoint Borel sets, then we represent the
disjunctive question \emph{``Is the particle in $S_{1}\cup S_{2}$?''}
formally by
\begin{equation} E^{e}(S_{1})\oplus E^{e}(S_{2})\df 
  E^{e}(S_{1})+E^{e}(S_{2}) . \label{first}
\end{equation}
The conjunctive question \emph{``Is the particle in $S_{1}$ and
  $S_{2}$?''} is represented by the operator $E^{e}(S_{1}\cap S_{2})$.
In general, however, it is not the case that
\begin{equation} E^{e}(S_{1}\cap S_{2})=E^{e}(S_{1}) E^{e}(S_{2})
  \label{conj}. \end{equation} In fact, the
operator on the right-hand side of Eq.~\ref{conj} will not generally
be in the range of $E^{e}$.  Finally, the question \emph{``Is the
  particle in the complement of $S$?''} is represented formally by
\begin{equation}
\neg E^{e}(S) \df E^{e} \left( \mathbb{R}\backslash S \right)
.\end{equation}
   
We obtain the standard projection valued (PV) measure $E$ if we apply
the above construction to the case in which we have absolute
confidence in the accuracy of our measuring apparatus.  Formally, if
we let $\delta$ be the Dirac delta function, then
\begin{equation}
E^{\delta}(S)= 
(\chi _{S}*\delta )(\widehat{Q}) =\chi _{S}(\widehat{Q})=E(S) ,\end{equation}
for all Borel sets $S$.  On the other hand, if $e$ has non-zero
deviation (i.e., $e$ is an integrable \emph{function}), then for
each Borel set $S$, $(\chi _{S}*e)$ is a uniformly continuous 
function~\cite[Prop.~3.2]{hirsch}.

\newcommand{\bi}{\mathbf{I}}

Realistically, then, the family $\alg{E}$ of experimental propositions
about the location of the particle should consist only of elements of
the form $E^{e}(S)$, where $e$ is some confidence function with
non-zero deviation.  Let's be generous, though, and suppose that
$\alg{E}$ contains all operators of the form $f(\widehat{Q})$ where
$f$ is a uniformly continuous function from $\mathbb{R}$ into $[0,1]$.
If we say that $A\perp B$ just in case $A+B\leq \mathbf{I}$, then we may extend
the exclusive disjunction defined in Eq.~\ref{first} by setting
\begin{equation}
A\oplus B \df A+B , \end{equation} when $A\perp B$.  It is easy to
verify then that $(\alg{E},\oplus ,\bz ,\mathbf{I})$ is an \emph{effect algebra}, 
or \emph{unsharp quantum logic}~\cite{chiara,chiara2,foulis}.  (Recall
that every Boolean algebra is an effect algebra if we set: $A\perp B$
iff. $A\wedge B=\bz$, and $A\oplus B=A\vee B$~\cite{foulis}.)  
Thus, in particular, for each $A\in \alg{E}$, there is a unique
element $\neg A=\bi -A\in \alg{E}$ such that $A\perp \neg A$ and $A\oplus
\neg A=\bi$.  For $A,B\in \alg{E}$, we say that $A\leq B$ just in case
there is some $C\perp A$ such that $A\oplus C=B$.  A \emph{state} on
$\alg{E}$ is a mapping $\omega :\alg{E}\mapsto [0,1]$ such that
$\omega (\mathbf{I})=1$ and \begin{equation}
 \omega (A\oplus B)=\omega (A)+\omega (B), \end{equation} whenever $A\oplus B$ is
defined.  From
this it follows that whenever $A,B\in \alg{E}$ and $A\leq B$, then
$\omega (A)\leq \omega (B)$ and \begin{equation} \omega (B-A)=\omega
  (B)-\omega (A) .\end{equation}  
We say that a state $\omega$ of $\alg{E}$ is pure just in case: If 
$\omega =a\rho +(1-a)\tau$, for some $a\in (0,1)$, then $\omega =\rho 
=\tau$.  Unlike the Boolean case, however, a pure state of a general
effect algebra may take on any value in the interval $[0,1]$.
(For example, even if Bob hits the bullseye on each try out
of $100$, our less than ideal score-keeper might only credit him with only,
say, $98$ hits.)  

I began this section with the idea that there are too many observables
in $\alg{O}_{Q}\cong \quo$, and that we should be more modest about
what can actually be measured.  Let me show precisely, now, how the
effect algebra $\alg{E}$ does represent a more modest perspective on
what the \emph{experimental} propositions about position truly are.

Both $\alg{E}$ and $\alg{O}_{Q}$ are subsets of the algebra
$\alg{R}_{Q}$ of all operators of the form $f(\widehat{Q})$, where $f$
is some Borel function from $\mathbb{R}$ into $\mathbb{C}$.  Although
the intersection of $\alg{E}$ and $\alg{O}_{Q}$ contains only $\bz$
and $\mathbf{I}$, we should think of $\alg{E}$ as containing a much
smaller set of observables than $\alg{O}_{Q}$.  Indeed, any element in
$\alg{E}$ may be uniformly approximated by linear combinations of
elements in $\alg{O}_{Q}$ (by the spectral theorem).  Thus, any pure
state $\omega$ on $\quo$ will give rise to a unique pure state $\omega
|_{\alg{E}}$ on $\alg{E}$.  (This slight abuse of notation is
justified by the fact that pure states on $\alg{O}_{Q}$ extend
uniquely to pure states of the von Neumann algebra $\alg{R}_{Q}$.)  On
the other hand, since the family of uniformly continous functions is
closed under linear combinations and uniform limits, no element of
$\quo$ (other than $\bz$ and $\mathbf{I}$) can be approximated by
linear combinations of elements in $\alg{E}$.  Thus, $\alg{E}$ is
``coarser grained'' than $\alg{O}_{Q}$ in the sense that two states
which give different outcomes for measurements in $\alg{O}_{Q}$ may
give identical outcomes for all measurements in $\alg{E}$.  And,
indeed, the elements of $\alg{E}$ do not permit us to distinguish
between states that correspond to a common point $\lambda \in
\mathbb{R}$.

\begin{prop}  Let $\omega ,\rho$ be pure states of $\alg{O}_{Q}$, both
  of which converge to $\lambda$.  Then,
  \begin{equation}
\omega (f(\widehat{Q}))=\rho (f(\widehat{Q}))=f(\lambda ) ,
\label{fec} \end{equation} for all $f(\widehat{Q})\in \alg{E}$.  
\end{prop}

\begin{proof} See Prop.~3.4 of Ref.~\ref{fink}.  \end{proof}

This proposition gives us everything we want.  For each $\lambda \in
\mathbb{R}$, there is precisely one (explicitly defined) pure state
$\omega _{\lambda}$ of $\alg{E}$ corresponding to $\lambda$.
(Moreover, it can be shown that every ``convergent'' state of
$\alg{E}$ is of this form.  See Proposition~\ref{one} in the
appendix.)  Therefore, any statistical state $\rho _{\psi}$ of
$\alg{E}$ decomposes as an integral
\begin{equation} \rho _{\psi}=\int _{\mathbb{R}}\omega _{\lambda} \,d\mu
  (\lambda ) ,\end{equation} which permits interpretation as a
measure of our ignorance of the precise, categorical location property
of individual particles in the ensemble.

\vspace{0.5em} {\it Acknowledgments:} I would like to thank E.
Schechter (Vanderbilt) for helpful correspondence.  Special thanks are
due to Rob Clifton for providing the impetus for this paper, and for
numerous discussions during its composition.  

\appendix
\section{Appendix}
\begin{lemma} Let $\omega$ be a state of $\alg{E}$.  Then
  $\omega (aA)=a\,\omega (A)$ for all $A\in \alg{E}$ and $a\in
  (0,1)$. \label{late} \end{lemma}
\begin{proof} Suppose first that $a=1/2$.  Then, $\omega (A)-
  \omega ((1/2)A)=\omega ((1/2)A)$ and so $\omega ((1/2)A)=(1/2)\omega
  (A)$.  It then follows easily by induction that $\omega (aA)=a\omega
  (A)$ when $a=2^{-n}$ for some $n\in \mathbb{N}$.  
  
  Suppose now that $a$ is a diadic rational; i.e., $a
  =m/2^{n}$ for some $m,n\in \mathbb{N}$, where $m< 2^{n}$.  Note then
  that \begin{equation} \frac{m}{2^{n}}A=\underbrace{\frac{1}{2^{n}}A\oplus
    \frac{1}{2^{n}}A\oplus \cdots \oplus \frac{1}{2^{n}}A}_{\mbox{$m$
  times}} 
    ,\end{equation} and so
\begin{equation} \omega \left( \frac{m}{2^{n}}A \right)
    =m\omega \left( \frac{1}{2^{n}}A \right) =\frac{m}{2^{n}}\omega
    (A) .\end{equation} 
Finally, let $a$ be an arbitrary element
  in $(0,1)$.  Then, $a =\lim _{n}a _{n}$ for some
  strictly decreasing sequence $\{ a _{n} \}\subseteq (0,1)$ of
  diadic rationals.  Now, for every $N\in \mathbb{N}$, there is some
  $M\in \mathbb{N}$ such that
\begin{equation} a _{n}A -aA \leq 2^{-N}\mathbf{I} ,\end{equation}
for all $n\geq M$, and therefore \begin{equation} \omega \left(
    a_{n}A-aA \right) \leq 2^{-N} ,\end{equation}
for all $n\geq M$.  We may also assume that $a_{n}-a\leq
2^{-N}$ when $n\geq M$.  Thus, \begin{eqnarray} \left| \omega (aA)
-a\,\omega (A) \right| &=&\left| \omega (aA)-\omega
    (a_{n}A) +a_{n}\omega (A)-a\,\omega (A) \right| \\
  &\leq& \left| \omega (aA)-\omega (a_{n}A) \right| +
  \left| a_{n}\omega (A)-a\,\omega (A) \right| \\
  &\leq &2^{-N}+2^{-N} .\end{eqnarray} Since this is true for all
$N\in \mathbb{N}$, it follows that $\omega (aA)=a\,\omega (A)$.
    \end{proof}

There are pure states on $\alg{E}$ that assign $0$ to all elements of
the form $E^{e}(S)$, where $S$ is a compact subset of $\mathbb{R}$.
(That such states exist can be seen from the proof of
Proposition~\ref{one} below.)  When $S$ is compact, then the function
$f\df(\chi _{S}*e)$ vanishes at infinity; i.e., for each $\epsilon
>0$, there in a $N\in \mathbb{N}$ such that $f(x)<\epsilon$ when
$\abs{x}>N$~\cite[Prop.~3.2]{hirsch}.  This motivates the following
definition.

\begin{defn} Let $\omega$ be a pure state of $\alg{E}$.  We say that
  $\omega$ converges just in case $\omega (f(\widehat{Q}))>0$ for some
  $f$ that vanishes at infinity.  \end{defn} 

\begin{prop}  Let $\omega$ be a convergent pure state of $\alg{E}$.
  Then, $\omega (f(\widehat{Q}))=f(\lambda )$ for some $\lambda \in
  \mathbb{R}$.  \label{one} \end{prop}

\begin{proof} Recall that the Stone-Cech compactification $\beta 
  \mathbb{R}$ of $\mathbb{R}$ is the unique compact Hausdorff space
  such that every bounded continuous function $f:\mathbb{R}\mapsto
  \mathbb{C}$ can be extended uniquely to a continuous function
  $\overline{f}:\beta \mathbb{R}\mapsto \mathbb{C}$. Let
  $BUC(\mathbb{R})$ denote the set of bounded, uniformly continuous
  functions from $\mathbb{R}$ into $\mathbb{C}$.  Since a uniform
  limit of elements in $BUC(\mathbb{R})$ is again in
  $BUC(\mathbb{R})$, it follows that there is a unique compact
  Hausdorff space $\gamma \mathbb{R}$ such that every $f\in
  BUC(\mathbb{R})$ has a unique extension to a continuous function
  $\overline{f}:\gamma \mathbb{R}\mapsto \mathbb{C}$.  We may refer to
  $\gamma \mathbb{R}$ as the uniform compactification of $\mathbb{R}$.
  Let $C(\gamma \mathbb{R} )$ denote the set of continuous functions
  from $\gamma \mathbb{R}$ into $\mathbb{C}$.  Thus, there is an
  algebraic isomorphism from $\buc$ onto $C(\gamma \mathbb{R})$ and we
  may identify $\alg{E}$ with the family of functions in $C(\gamma
  \mathbb{R})$ with range $[0,1]$.
  
  We show that every (pure) state $\omega$ of $\alg{E}$ has a unique
  extension to a (pure) state $\widehat{\omega}$ of the
  $C^{*}$-algebra $C(\comp )$.  We may then appeal to the result that
  any pure state $\widehat{\omega}$ on $C(\comp )$ is of the form
  $\widehat{\omega}(f)=f(x_{0})$ for some $x_{0}\in
  \comp$~\cite[Corollary~3.4.2]{kr}.
  
  Let $\omega$ be a state of $\alg{E}$.  Let $C^{+}(\gamma
  \mathbb{R})$ denote the set of functions from $\gamma \mathbb{R}$
  into $[0,+\infty )$.  If $f\in C^{+}(\gamma \mathbb{R})$, then
  $f/\norm{f}\in \alg{E}$ and we may define
\begin{equation} \widehat{\omega}(f)\df \norm{f}\cdot \omega \left( f/\norm{f}
  \right) .\end{equation}
Using Lemma~\ref{late}, it is easy to see that $\widehat{\omega}$ is
homogenous with respect to positive real numbers, and it is
straightforward to verify that $\widehat{\omega}$ is additive over
positive functions.  Thus, $\widehat{\omega}$ is an additive function
from $C^{+}(\gamma \mathbb{R})$ into $[0,\infty )$.  It follows then that
$\widehat{\omega}$ extends uniquely to a linear mapping from $C(\gamma
\mathbb{R})$ into $\mathbb{C}$~\cite[Prop. 11.55]{schec}.    

Suppose now that $\omega$ is a pure state of $\alg{E}$, and let
$\widehat{\omega}$ be the unique extension of $\omega$ to $C(\gamma
\mathbb{R})$ as defined above.  Suppose that $\widehat{\omega}=a\rho
+(1-a)\tau$, where $\rho ,\tau$ are states of $C(\gamma \mathbb{R})$
and $a\in (0,1)$.  Then $\rho \,|_{\alg{E}}$ and $\tau \,|_{\alg{E}}$
are states of $\alg{E}$, and $\omega =a\rho \,|_{\alg{E}}+(1-a)\tau
\,|_{\alg{E}}$.  However, since $\omega$ is a pure state of
$\alg{E}$, we have $\omega =\rho \,|_{\alg{E}}=\tau \,|_{\alg{E}}$ and
since extensions are unique, it follows that
$\widehat{\omega}=\rho=\tau$.  Thus, $\widehat{\omega}$ is a pure
state of $C(\gamma \mathbb{R})$.
  
Thus, there is some $x_{0}\in \comp$ such that
$\widehat{\omega}(f)=f(x_{0})$ for each $f\in C(\comp )$.  We show
that if $x_{0}\in \comp \backslash \mathbb{R}$, then the state
$\omega$ does not converge.  Indeed, since $\mathbb{R}$ is dense in
$\comp$, there is a net $(x_{a})_{a\in \mathbb{A}}$ in $\mathbb{R}$
such that $x_{a}\rightarrow x_{0}$.  Let $f\in \alg{E}$ be a function
that vanishes at infinity.  We show that $\overline{f}(x_{0})=0$.  Let
$\epsilon >0$ be given.  Then, there is an $N\in \mathbb{N}$ such that
$\overline{f}(x)<\epsilon$ for all $x\in \mathbb{R}$ with $\abs{x}\geq
N$.  However, since the set $[-N,N]$ is compact in $\gamma \mathbb{R}$
and $x_{0}\not\in \mathbb{R}$, there is some $b\in \mathbb{A}$ such
that $x_{a}\not\in [-N,N]$ for all $a\geq b$.  Since $f(x_{0})=\lim
f(x_{a})$, $f(x_{0})<\epsilon $.  Since this is true for any $\epsilon
>0$, it follows that $f(x_{0})=0$ and $\omega (f)=0$.  \end{proof}

\end{document}